%
%
%
%
\def\unredoffs{} \def\redoffs{\voffset=-.40truein\hoffset=-.40truein}
\def\speclscape{}
%
%
%
%
\newbox\leftpage \newdimen\fullhsize \newdimen\hstitle \newdimen\hsbody
\tolerance=1000\hfuzz=2pt
\catcode`\@=11 
\def\bigans{b }
\def\answ{b }
\ifx\answ\bigans\message{(This will come out unreduced.}
\magnification=1200\unredoffs\baselineskip=16pt plus 2pt minus 1pt
\hsbody=\hsize \hstitle=\hsize 
\else\message{(This will be reduced.} \let\l@r=L
\magnification=1000\baselineskip=16pt plus 2pt minus 1pt
\vsize=7truein \redoffs
\hstitle=8truein\hsbody=4.75truein\fullhsize=10truein\hsize=\hsbody
\output={\ifnum\pageno=0 
    \shipout\vbox{\speclscape{\hsize\fullhsize\makeheadline}
      \hbox to \fullhsize{\hfill\pagebody\hfill}}\advancepageno
    \else
    \almostshipout{\leftline{\vbox{\pagebody\makefootline}}}\advancepageno
    \fi}
\def\almostshipout#1{\if L\l@r \count1=1 \message{[\the\count0.\the\count1]}
        \global\setbox\leftpage=#1 \global\let\l@r=R
   \else \count1=2
    \shipout\vbox{\speclscape{\hsize\fullhsize\makeheadline}
        \hbox to\fullhsize{\box\leftpage\hfil#1}}  \global\let\l@r=L\fi}
\fi
%
\newcount\yearltd\yearltd=\year\advance\yearltd by -1900

\def\Title#1#2{\nopagenumbers\abstractfont\hsize=\hstitle\rightline{#1}%
\vskip 1in\centerline{\titlefont #2}\abstractfont\vskip
.5in\pageno=0}
%
%

\def\draftmode{\message{ DRAFTMODE }\def\draftdate{{\rm preliminary draft:
\number\month/\number\day/\number\yearltd\ \ \hourmin}}%
\headline={\hfil\draftdate}\writelabels\baselineskip=20pt plus 2pt
minus 2pt
   {\count255=\time\divide\count255 by 60 \xdef\hourmin{\number\count255}
    \multiply\count255 by-60\advance\count255 by\time
    \xdef\hourmin{\hourmin:\ifnum\count255<10 0\fi\the\count255}}}
\def\nolabels{\def\wrlabeL##1{}\def\eqlabeL##1{}\def\reflabeL##1{}}
\def\writelabels{\def\wrlabeL##1{\leavevmode\vadjust{\rlap{\smash%
{\line{{\escapechar=` \hfill\rlap{\sevenrm\hskip.03in\string##1}}}}}}}%
\def\eqlabeL##1{{\escapechar-1\rlap{\sevenrm\hskip.05in\string##1}}}%
\def\reflabeL##1{\noexpand\llap{\noexpand\sevenrm\string\string\string##1}}}
\nolabels
%
\global\newcount\secno \global\secno=0 \global\newcount\meqno
\global\meqno=1
\def\newsec#1{\global\advance\secno by1\message{(\the\secno. #1)}
\global\subsecno=0\eqnres@t\noindent{\bf\the\secno. #1}
\writetoca{{\secsym} {#1}}\par\nobreak\medskip\nobreak}
\def\eqnres@t{\xdef\secsym{\the\secno.}\global\meqno=1\bigbreak\bigskip}
\def\sequentialequations{\def\eqnres@t{\bigbreak}}\xdef\secsym{}
\global\newcount\subsecno \global\subsecno=0
\def\subsec#1{\global\advance\subsecno by1\message{(\secsym\the\subsecno. #1)}
\ifnum\lastpenalty>9000\else\bigbreak\fi
\noindent{\it\secsym\the\subsecno. #1}\writetoca{\string\quad
{\secsym\the\subsecno.} {#1}}\par\nobreak\medskip\nobreak}
\def\appendix#1#2{\global\meqno=1\global\subsecno=0\xdef\secsym{\hbox{#1.}}
\bigbreak\bigskip\noindent{\bf Appendix #1. #2}\message{(#1. #2)}
\writetoca{Appendix {#1.} {#2}}\par\nobreak\medskip\nobreak}
%
%
\def\eqnn#1{\xdef #1{(\secsym\the\meqno)}\writedef{#1\leftbracket#1}%
\global\advance\meqno by1\wrlabeL#1}
\def\eqna#1{\xdef #1##1{\hbox{$(\secsym\the\meqno##1)$}}
\writedef{#1\numbersign1\leftbracket#1{\numbersign1}}%
\global\advance\meqno by1\wrlabeL{#1$\{\}$}}
\def\eqn#1#2{\xdef #1{(\secsym\the\meqno)}\writedef{#1\leftbracket#1}%
\global\advance\meqno by1$$#2\eqno#1\eqlabeL#1$$}
%
\newskip\footskip\footskip14pt plus 1pt minus 1pt 
\def\footnotefont{\ninepoint}\def\f@t#1{\footnotefont #1\@foot}
\def\f@@t{\baselineskip\footskip\bgroup\footnotefont\aftergroup\@foot\let\next}
\setbox\strutbox=\hbox{\vrule height9.5pt depth4.5pt width0pt}
\global\newcount\ftno \global\ftno=0
\def\foot{\global\advance\ftno by1\footnote{$^{\the\ftno}$}}
%
\newwrite\ftfile
\def\footend{\def\foot{\global\advance\ftno by1\chardef\wfile=\ftfile
$^{\the\ftno}$\ifnum\ftno=1\immediate\openout\ftfile=foots.tmp\fi%
\immediate\write\ftfile{\noexpand\smallskip%
\noexpand\item{f\the\ftno:\ }\pctsign}\findarg}%
\def\footatend{\vfill\eject\immediate\closeout\ftfile{\parindent=20pt
\centerline{\bf Footnotes}\nobreak\bigskip\input foots.tmp }}}
\def\footatend{}
%
%
\global\newcount\refno \global\refno=1
\newwrite\rfile
\def\ref{[\the\refno]\nref}
\def\nref#1{\xdef#1{[\the\refno]}\writedef{#1\leftbracket#1}%
\ifnum\refno=1\immediate\openout\rfile=refs.tmp\fi
\global\advance\refno by1\chardef\wfile=\rfile\immediate
\write\rfile{\noexpand\item{#1\
}\reflabeL{#1\hskip.31in}\pctsign}\findarg}
\def\findarg#1#{\begingroup\obeylines\newlinechar=`\^^M\pass@rg}
{\obeylines\gdef\pass@rg#1{\writ@line\relax #1^^M\hbox{}^^M}%
\gdef\writ@line#1^^M{\expandafter\toks0\expandafter{\striprel@x #1}%
\edef\next{\the\toks0}\ifx\next\em@rk\let\next=\endgroup\else\ifx\next\empty%
\else\immediate\write\wfile{\the\toks0}\fi\let\next=\writ@line\fi\next\relax}}
\def\striprel@x#1{} \def\em@rk{\hbox{}}
\def\lref{\begingroup\obeylines\lr@f}
\def\lr@f#1#2{\gdef#1{\ref#1{#2}}\endgroup\unskip}
\def\semi{;\hfil\break}
\def\addref#1{\immediate\write\rfile{\noexpand\item{}#1}} 
\def\startrefs#1{\immediate\openout\rfile=refs.tmp\refno=#1}
\def\xref{\expandafter\xr@f}\def\xr@f[#1]{#1}
\def\refs#1{\count255=1[\r@fs #1{\hbox{}}]}
\def\r@fs#1{\ifx\und@fined#1\message{reflabel \string#1 is undefined.}%
\nref#1{need to supply reference \string#1.}\fi%
\vphantom{\hphantom{#1}}\edef\next{#1}\ifx\next\em@rk\def\next{}%
\else\ifx\next#1\ifodd\count255\relax\xref#1\count255=0\fi%
\else#1\count255=1\fi\let\next=\r@fs\fi\next}
%

%
\newwrite\ffile\global\newcount\figno \global\figno=1
\def\fig{fig.~\the\figno\nfig}
\def\nfig#1{\xdef#1{fig.~\the\figno}%
\writedef{#1\leftbracket fig.\noexpand~\the\figno}%
\ifnum\figno=1\immediate\openout\ffile=figs.tmp\fi\chardef\wfile=\ffile%
\immediate\write\ffile{\noexpand\medskip\noexpand\item{Fig.\
\the\figno. }
\reflabeL{#1\hskip.55in}\pctsign}\global\advance\figno
by1\findarg}
\def\xfig{\expandafter\xf@g}\def\xf@g fig.\penalty\@M\ {}
\def\figs#1{figs.~\f@gs #1{\hbox{}}}
\def\f@gs#1{\edef\next{#1}\ifx\next\em@rk\def\next{}\else
\ifx\next#1\xfig #1\else#1\fi\let\next=\f@gs\fi\next}
\newwrite\lfile
{\escapechar-1\xdef\pctsign{\string\%}\xdef\leftbracket{\string\{}
\xdef\rightbracket{\string\}}\xdef\numbersign{\string\#}}

\def\writestop{\def\writestoppt{\immediate\write\lfile{\string\pageno%
\the\pageno\string\startrefs\leftbracket\the\refno\rightbracket%
\string\def\string\secsym\leftbracket\secsym\rightbracket%
\string\secno\the\secno\string\meqno\the\meqno}\immediate\closeout\lfile}}
\def\writestoppt{}\def\writedef#1{}
\def\seclab#1{\xdef #1{\the\secno}\writedef{#1\leftbracket#1}\wrlabeL{#1=#1}}
\def\subseclab#1{\xdef #1{\secsym\the\subsecno}%
\writedef{#1\leftbracket#1}\wrlabeL{#1=#1}}
\newwrite\tfile \def\writetoca#1{}
\def\leaderfill{\leaders\hbox to 1em{\hss.\hss}\hfill}
\def\writetoc{\immediate\openout\tfile=toc.tmp
     \def\writetoca##1{{\edef\next{\write\tfile{\noindent ##1
     \string\leaderfill {\noexpand\number\pageno} \par}}\next}}}

%
\catcode`\@=12 
%
\edef\tfontsize{\ifx\answ\bigans scaled\magstep3\else
scaled\magstep4\fi} \font\titlerm=cmr10 \tfontsize
\font\titlerms=cmr7 \tfontsize \font\titlermss=cmr5 \tfontsize
\font\titlei=cmmi10 \tfontsize \font\titleis=cmmi7 \tfontsize
\font\titleiss=cmmi5 \tfontsize \font\titlesy=cmsy10 \tfontsize
\font\titlesys=cmsy7 \tfontsize \font\titlesyss=cmsy5 \tfontsize
\font\titleit=cmti10 \tfontsize \skewchar\titlei='177
\skewchar\titleis='177 \skewchar\titleiss='177
\skewchar\titlesy='60 \skewchar\titlesys='60
\skewchar\titlesyss='60
\def\titlefont{\def\rm{\fam0\titlerm}
\textfont0=\titlerm \scriptfont0=\titlerms
\scriptscriptfont0=\titlermss \textfont1=\titlei
\scriptfont1=\titleis \scriptscriptfont1=\titleiss
\textfont2=\titlesy \scriptfont2=\titlesys
\scriptscriptfont2=\titlesyss \textfont\itfam=\titleit
\def\it{\fam\itfam\titleit}\rm}
 \ifx\answ\bigans\else scaled\magstep1\fi
\ifx\answ\bigans\def\abstractfont{\tenpoint}\else
\font\abssl=cmsl10 scaled \magstep1 \font\absrm=cmr10
scaled\magstep1 \font\absrms=cmr7 scaled\magstep1
\font\absrmss=cmr5 scaled\magstep1 \font\absi=cmmi10
scaled\magstep1 \font\absis=cmmi7 scaled\magstep1
\font\absiss=cmmi5 scaled\magstep1 \font\abssy=cmsy10
scaled\magstep1 \font\abssys=cmsy7 scaled\magstep1
\font\abssyss=cmsy5 scaled\magstep1 \font\absbf=cmbx10
scaled\magstep1 \skewchar\absi='177 \skewchar\absis='177
\skewchar\absiss='177 \skewchar\abssy='60 \skewchar\abssys='60
\skewchar\abssyss='60
\def\abstractfont{\def\rm{\fam0\absrm}
\textfont0=\absrm \scriptfont0=\absrms \scriptscriptfont0=\absrmss
\textfont1=\absi \scriptfont1=\absis \scriptscriptfont1=\absiss
\textfont2=\abssy \scriptfont2=\abssys \scriptscriptfont2=\abssyss
\textfont\itfam=\bigit \def\it{\fam\itfam\bigit}\def\footnotefont{\tenpoint}%
\textfont\slfam=\abssl \def\sl{\fam\slfam\abssl}%
\textfont\bffam=\absbf \def\bf{\fam\bffam\absbf}\rm}\fi
\def\tenpoint{\def\rm{\fam0\tenrm}
\textfont0=\tenrm \scriptfont0=\sevenrm \scriptscriptfont0=\fiverm
\textfont1=\teni  \scriptfont1=\seveni  \scriptscriptfont1=\fivei
\textfont2=\tensy \scriptfont2=\sevensy \scriptscriptfont2=\fivesy
\textfont\itfam=\tenit \def\it{\fam\itfam\tenit}\def\footnotefont{\ninepoint}%
\textfont\bffam=\tenbf
\def\bf{\fam\bffam\tenbf}\def\sl{\fam\slfam\tensl}\rm}
\font\ninerm=cmr9 \font\sixrm=cmr6 \font\ninei=cmmi9
\font\sixi=cmmi6 \font\ninesy=cmsy9 \font\sixsy=cmsy6
\font\ninebf=cmbx9 \font\nineit=cmti9 \font\ninesl=cmsl9
\skewchar\ninei='177 \skewchar\sixi='177 \skewchar\ninesy='60
\skewchar\sixsy='60
\def\ninepoint{\def\rm{\fam0\ninerm}
\textfont0=\ninerm \scriptfont0=\sixrm \scriptscriptfont0=\fiverm
\textfont1=\ninei \scriptfont1=\sixi \scriptscriptfont1=\fivei
\textfont2=\ninesy \scriptfont2=\sixsy \scriptscriptfont2=\fivesy
\textfont\itfam=\ninei \def\it{\fam\itfam\nineit}\def\sl{\fam\slfam\ninesl}%
\textfont\bffam=\ninebf \def\bf{\fam\bffam\ninebf}\rm}
%
%

\hyphenation{anom-aly anom-alies coun-ter-term coun-ter-terms}
\def\inv{^{\raise.15ex\hbox{${\scriptscriptstyle -}$}\kern-.05em 1}}

\def\Dsl{\,\raise.15ex\hbox{/}\mkern-13.5mu D} 
\def\dsl{\raise.15ex\hbox{/}\kern-.57em\partial}

 \def\Tr{{\rm Tr}}
\font\bigit=cmti10 scaled \magstep1
\def\lspace{\ifx\answ\bigans{}\else\qquad\fi}
\def\lbspace{\ifx\answ\bigans{}\else\hskip-.2in\fi} 
\def\boxeqn#1{\vcenter{\vbox{\hrule\hbox{\vrule\kern3pt\vbox{\kern3pt
      \hbox{${\displaystyle #1}$}\kern3pt}\kern3pt\vrule}\hrule}}}
\def\mbox#1#2{\vcenter{\hrule \hbox{\vrule height#2in
          \kern#1in \vrule} \hrule}}  
%
\def\CA{{\cal A}}   
 \def\CH{{\cal H}}

\def\darr#1{\raise1.5ex\hbox{$\leftrightarrow$}\mkern-16.5mu #1}

\def\half{{\textstyle{1\over2}}} 
\def\roughly#1{\raise.3ex\hbox{$#1$\kern-.75em\lower1ex\hbox{$\sim$}}}

\def\pl#1#2#3{Phys. Lett. {\bf #1B} (#2) #3}

\def\jhep#1#2#3{JHEP {\bf#1}(#2) #3}

\def\IB{\relax\hbox{$\inbar\kern-.3em{\rm B}$}}
\def\IC{{\bf C}}
\def\ID{\relax\hbox{$\inbar\kern-.3em{\rm D}$}}
\def\IE{\relax\hbox{$\inbar\kern-.3em{\rm E}$}}
\def\IF{\relax\hbox{$\inbar\kern-.3em{\rm F}$}}
\def\IG{\relax\hbox{$\inbar\kern-.3em{\rm G}$}}
\def\IGa{\relax\hbox{${\rm I}\kern-.18em\Gamma$}}
\def\IH{\relax{\rm I\kern-.18em H}}
\def\IK{\relax{\rm I\kern-.18em K}}
\def\IL{\relax{\rm I\kern-.18em L}}
\def\IP{\relax{\rm I\kern-.18em P}}
\def\IR{{\bf R}}
\def\IZ{{\bf Z}}
\def\II{\relax{\rm I\kern-.18em I}}

\def\ndt{{\noindent}}


\def\CA{{\cal A}}

\def\CH{{\cal H}}

\def\CM{{\cal M}}

\def\CZ{{\cal Z}}

\def\p{\partial}


\def\wb{\bar{w}}
\def\sb{\bar{s}}

\def\zb{\bar{z}}

\def\Tr{{\rm Tr}}


\def\inbar{\,\vrule height1.5ex width.4pt depth0pt}

\def\a{{\alpha}}

\def\d{{\delta}}

\def\vf{{\varphi}}
\def\m{{\mu}}

\def\l{{\lambda}}
\def\s{{\sigma}}
\def\t{{\tau}}
\def\th{{\theta}}
\def\vt{{\vartheta}}

\def\lref{\begingroup\obeylines\lr@f}
\def\lr@f#1#2{\gdef#1{\ref#1{#2}}\endgroup\unskip}

\lref\douglasmoore{M.~Douglas, G.~Moore, ``D-branes, Quivers, and
ALE Instantons'', hep-th/9603167}

\lref\gsw{M.~Green, J.~Schwarz, E.~Witten, ``Superstring theory'', Cambridge Univeristy Press, 1987}

\lref\bcbb{J.~Khoury, B.~Ovrut, N.~Seiberg, P.~Steinhart, N.~Turok, ``From Big Crunch to Big Bang'', hep-th/0108187}

\lref\joe{J.~Polchinski, ``String Theory'', Cambridge University Press, 1998}

\lref\horsteif{G.~Horowitz, A.~Steif, \pl{258}{1991}{91}}

\lref\cornalba{L.~Cornalba, M.~Costa, hep-th/0203031}

\lref\vijay{V.~Balasubramanian, S.~F.~Hassan, E.~Keski-Vakkuri,
A.~Naqvi, hep-th/0202187}

\lref\sb{M.~Gutperle, A.~Strominger, hep-th/0202210}

\lref\polyakov{A.~Polyakov, ``Gauge fields and space-time'', hep-th/0110196}

\lref\seibpol{J.~Polchinski, N.~Seiberg, unpublished}

\lref\bachasporrati{C.~Bachas, M.~Porrati, ``Pair creation of open
strings in an electric field'', hep-th/9209032,
\pl{296}{1992}{77-84}}

\lref\costas{C.~Bachas, N.~Nekrasov, to appear}

\lref\aps{A.~Adams, J.~Polchinski, E.~Silverstein, ``Don't Panic!
Closed String Tachyons in ALE Spacetimes'', hep-th/0108075,
\jhep{0110}{2001}{029} }

\lref\hmk{C.~Vafa, hep-th/0111051\semi J.~Harvey, D.~Kutasov,
E.~Martinec, G.~Moore, hep-th/0111154\semi A.~Dabholkar, C.~Vafa,
hep-th/0111155}

\def\ihes{{\it Institute des Hautes Etudes Scientifiques, Bures-sur-Yvette, F-91440 France}}
\def\itep{{ Institute for Theoretical and
Experimental Physics, 117259 Moscow, Russia}}

\Title{\vbox{\baselineskip 10pt \hbox{IHES-P/02/13}
\hbox{ITEP-TH-14/02} \hbox{hep-th/0203112} {\hbox{ }}}}
{\vbox{\vskip -30 true pt \centerline{Milne Universe, Tachyons,
and Quantum Group} \centerline{}
\smallskip
\smallskip
\smallskip
\smallskip
\smallskip
\medskip
\vskip4pt }} \vskip -20 true pt \centerline{ Nikita
A.~Nekrasov\foot{On leave of absence from \itep}}
\smallskip\smallskip
\centerline{\ihes}  \smallskip \centerline{\tt e-mail:
nikita@ihes.fr}
\bigskip
We analyze the spectrum of the bosonic and superstring on the
orbifold of the space-time by a boost, leading to the cosmological
singularity. We show that the modular invariance leads to the
spectrum where the twisted sector tachyon, together with all other
twisted sector fields, present in the Euclidean version of the
orbifold, is absent. This makes impossible to resolve the
singularity by a marginal deformation of the worldsheet CFT. We
also establish a relation between the resolution of rotational
orbifolds in Euclidean and Lorentzian setups, and quantum groups.
The analysis confirms the impossibility of resolving the
cosmological orbifold singularity.

The article is based on two talks, given at ``Workshop on
ekpyrotic universe'', Annecy, 12/01, and at the XXX ITEP Winter
school, Moscow, 02/01.

\newsec{The model: Minkowskian orbifold}

It was proposed in \bcbb\ to model the transition between a
contracting and expanding phases of the Universe by an M-theory
background of the form ${\IR}^{9} \times {\CM}$, where ${\CM}$ is
the so-called two dimensional Milne universe, the space with
conical singularity: \eqn\mln{ds^2 = - dt^2 + {\l} ^2 t^2 dw^2}
where $w$ lives on a circle of the circumference ${2\pi}$ and
${\l} > 0$ is a parameter. If we interpret $w$ as the eleventh
dimensional circle, then from the ten dimensional perspective we
have a Universe with the line element: \eqn\tnd{ds^2_{10} = a^2
(t) \left( - dt^2 + d{\vec x}^2 \right), \qquad a (t) \propto
\vert t \vert^{1\over 8}} and the dilaton \eqn\dlt{{\phi} (t) =
{\phi} (1) \pm \sqrt{2 \cdot 9 \over 8} {\log} \vert t \vert}
which passes through a zero string coupling point for the $+1$
choice of sign at $t=0$. However the coupling is strong far away
from the cosmic singularity. To study this configuration using
string theory one is naturally led to compactifying one of the
flat nine dimensions, then to do the nine-ten flip and interpret
the same background as the strong coupling limit of the type IIA
string propagating on the space-time of the form ${\CM} \times
{\IR}^8$. The problem with the space ${\CM}$ is that the curvature
has a delta-function singularity at $t=0$. Therefore it fails to
define a worldsheet conformal theory. Nevertheless, the manifold
${\CM}$ is an orbifold \horsteif. Take the Minkowski space
${\IR}^{1,1}$, with the light-cone coordinates $x^{\pm}$:
\eqn\lc{x^{\pm} = {1\over \sqrt{2}} \left( x^0 \pm x^9 \right)}
The manifold ${\CM}$ is the quotient of the interior of the past-
and future- light-cones $x^{+} x^{-} > 0$ by the action of the
group  ${\IZ}$ generated by a boost: \eqn\bst{x^{\pm} \mapsto
e^{\pm 2\pi {\l}} x^{\pm}} together with the point $x^{\pm} = 0$.
It is not easy to confine the string to propagate in the interiour
of the light-cone, so instead we consider the orbifold of
${\IR}^{1,1}$ by the same action ${\IZ}$. The resulting space is
non-Hausdorff (because of the light-cone) but the string
propagation on it should be fine. Notice that the quotient $M$
contains four cones, from the regions where $x^{\pm}$ have some
definite signs. In particular the domains with $x^{+} x^{-} < 0$
give rise to the space ${\CM}^{\vee}$ with the metric
\eqn\antimilne{ds^2 = dt^2 - {\l}^2 t^2 dw^2} which contains
closed time-like loops.

\subsec{Particle in the orbifolded universe}

Before studying the properties of a string, let us start with
considering a relativistic particle of mass $m$ propagating in
${\IR}^{1,1}/{\IZ}$. The wavefunction of this particle must be a
${\IZ}$-invariant solution of Klein-Gordon equation on
${\IR}^{1,1}$. Choose any on-shell momentum $p = (p^{+}, p^{-})$
on the covering space: $2 p^{+} p^{-} = m^2$. The function:
\eqn\inf{{\Psi} (x)_{p; l} = \int_{\IR} \ {\rm d}w \ e^{i \left(
p^{+} x^{-} e^{-{\l}w}  +  p^{-} x^{+} e^{ {\l} w} + l w \right)},
\qquad l \in {\IZ} }is clearly a ${\IZ}$-invariant on-shell
function. Moreover, changing $(p^{+}, p^{-})$ to $(z\ p^{+},
z^{-1} \ p^{-})$, for $z
> 0$, simply multiplies \inf\ by an irrelevant
phase. By this transformation one can bring $p$ to
${1\over\sqrt{2}} (m,m)$ where $m$ can be positive or negative.
Let us introduce the appropriately normalized functions:
\eqn\fns{{\Phi}_{m,l} (x) = \sqrt{2 \vert m \vert} \
{\Psi}_{{1\over\sqrt{2}} (m,m) ; l} (x)} The functions \fns\ are
orthonormal: \eqn\ortn{( {\Phi}_{m,l}, {\Phi}_{m^{\prime},
l^{\prime}}) = \int_{{\IR}^{1,1}/{\IZ}} {\rm d}x^{+} {\rm d}x^{-}
\ {\Phi}_{m,l}^{*} (x) {\Phi}_{m^{\prime}, l^{\prime}}(x) =
{\d}_{l,l^{\prime}} \ {\d} (m - m^{\prime})} It is instructive to
look at the behaviour of the wavefunctions ${\Phi}_{m,l}$ in the
``bad'' space-time region, i.e. for $x^{+} x^{-} < 0$. We claim
that the motion of the massive particle in this region is finite.
Indeed, for sufficiently large negative $x^{+}x^{-}$ the function
${\Phi}_{m,l} (x)$ is exponentially decaying: \eqn\ed{{\Phi}_{m,l}
\sim {\exp} ( -  \vert m \vert \sqrt{ - 2 x^{+} x^{-}} ) , \qquad
- x^{+} x^{-} \gg {l^2 \over 2 {\l} m^2}}

\subsec{Bosonic string}

We proceed with consideration of the bosonic string propagating on
$M \times {\IR}^{24}$. The orbifold with respect to the group
\bst\ contains  twisted sectors, labeled by an integer $M$:
\eqn\tws{X^{\pm} ({\s} + {\pi} ) = e^{\pm 2\pi M {\l} } X^{\pm}
({\s})} We start with the $M=1$ sector. The subsequent formulae
are easily generalized to the general $M$ case, by the
substitution: ${\l} \to {\l} M$. The mode expansion of the
$X^{\pm}$ coordinates in the twisted sector is given by:
\eqn\mde{\eqalign{& X^{\pm}_{L} = \sum_{n} {{\tilde\a}^{\pm}_{n}
\over 2(\pm {\l} - in)} e^{2(\pm {\l} - in) ( t + {\s})} \cr &
X^{\pm}_{R} = - \sum_{n} {{\a}^{\pm}_{n} \over 2({\pm\l} + in)}
e^{-2({\pm\l} + in) (t - {\s})} \cr}  }

Straightforward quantization leads to the commutation relations:
\eqn\cmtr{\eqalign{ & [ {\a}_{n}^{+}, {\a}_{m}^{-} ] = {\d}_{m+n}
( m + i{\l}) \cr & [ {\tilde\a}_{n}^{+}, {\tilde\a}_{m}^{-} ] =
{\d}_{m+n} (m - i{\l}) \cr}} Note that the oscillators with
$n=m=0$ form a Heisenberg algebra. The Virasoro generators $L_{n}$
are calculated from the classical stress-energy tensor and then
normal ordered with the result: \eqn\virs{\eqalign{& L_{m} = -
\sum_n {\a}^{+}_{n} {\a}^{-}_{m-n} + \sum_n {\a}^i_{n}
{\a}^{i}_{m-n}, \qquad m \neq 0 \cr & L_0 = {{\l}^2 \over 2} -
{\half} \left( {\a}^+_0 {\a}^{-}_0 + {\a}^{-}_{0} {\a}^{+}_{0}
\right) + {\half} {\vec p}^2 + \sum_{n > 0} {\vec \a}_{-n}
{\vec\a}_n - {\a}^{+}_{-n} {\a}^{-}_n - {\a}^{-}_{-n} {\a}^{+}_{n}
\cr}}where ${\vec\a}_n$ are the transverse oscillators. The
generators \virs\ form the standard algebra $$ [L_n, L_m] = (n-m)
L_{n+m} + {26 \over 12} (n^3 - n) {\d}_{n+m} $$ The ground state
is annihilated by the positive frequency oscillators:
\eqn\gnds{{\a}^{\m}_{n} \vert 0 \rangle_{R} = 0 , \qquad
{\tilde\a}^{\m}_{n} \vert 0 \rangle_{L} = 0, \qquad n
> 0} The zero modes ${\a}_0^{\pm}$ have to be quantized with some
care.

\subsec{One-loop partition function}

In this section we shall find that the correct way to quantize the
zero modes ${\a}^{\pm}_0$ is by analytic continuation of the
ordinary creation-annihilation operators of the harmonic
oscillator to the case of imaginary frequency (just as in the open
string case \bachasporrati) .

Consider the one-loop string measure. It splits as a sum over all
winding sectors, which we shall label by two integers $(L_1,
L_2)$. In the $(L_1, L_2)$ sector the boundary conditions on the
worldsheet field $X^{\pm}$ are: \eqn\bndryc{X^{\pm} ( z + m + n
{\tau}) = {\exp} {\pm 2\pi\l} \left( L_1 \ m + L_2 \ n \right) \
X^{\pm} (z)} where we work on the Euclidean torus with the complex
structure specified by the modular parameter ${\tau} = {\tau}_1 +
i {\tau}_2 $. We might also have to use the Lorentzian tori, for
which ${\tau}$ and ${\bar\tau}$ are both real and ${\tau} -
{\bar\tau} > 0$. In each case the metric on the torus is given by:
\eqn\mtr{ds^2 = (dx + {\tau} dy ) (dx + {\bar\tau} dy)} where
$(x,y) \sim (x + m, y + n)$, $m,n \in{\IZ}$. There are several
ways of computing ${\CZ}_{(L_1, L_2)} ({\t})$. One can directly
calculate the regularized determinant of the Laplace operator, or
one can calculate ${\CZ}_{(0,L)}$ and then use modular group to
generate the rest. The full partition function \eqn\flprt{{\CZ}
({\tau}, {\bar\tau}) = \sum_{(L_1, L_2) \in {\IZ}^2} \
{\CZ}_{(L_1, L_2)} } should be modular invariant. It is clear that
$(L_1, L_2)$ transforms as a doublet of $SL_2 ({\IZ})$. We expect
that \eqn\mdlrtr{{\CZ}_{(a L_1 - c L_2, - b L_1 + d L_2)} \left({
a {\tau} + b \over c {\tau} + d}\right) = {\CZ}_{(L_1, L_2)}
({\tau})} for $ad - bc = 1$. We can split the sum over $(L_1,
L_2)$ as a sum over the $N = gcd(L_1, L_2)$ and then for fixed $N
\neq 0$ the sum over the mutually prime integers $(p,q)$, s.t.
$L_1 = N p , L_2 N q)$. The latter sum is the sum over the coset
space $SL_2 ({\IZ})/{\IZ}$. In the subsequent integration over the
modular domain \eqn\olit{\int_{{\CM}_{1,1}} {d{\tau} d{\bar\tau}
\over {\tau}_2^2 }  \ {\CZ} ({\tau}, {\bar\tau})}this summation
unfolds the moduli space ${\CM}_{1,1}$ and makes it into a strip
$\vert \tau_1 \vert \leq {1\over 2}, {\tau}_2 > 0$.

For $N=0$ the boundary conditions are periodic and one gets the
same answer as in the ordinary bosonic string case:
\eqn\zrsctr{{\CZ}_{(0,0)} (\tau) = {\rm Vol} ({\CM} \times
{\IR}^{24}) {1\over{{\tau}_2^{12} {\vert \eta (\tau)
\vert^{48}}}}} with the only difference being the reduction of the
volume of the target space. For $N\neq 0$ the partition function
may or may not contain the divergent factor coming from the zero
modes, which are always present for $N=0$. The $24$ transverse
coordinates will of course contribute the factor of ${\rm
Vol}({\IR}^{24})$ independently of $N, {\l}$.

Let us now analyze the conditions on the zero modes coming from
the ${\IR}^{1,1}/{\IZ}$ part. The eigenvalues of the Laplace
operator on the functions obeying \bndryc, for the metric \mtr,
are: \eqn\eigenf{{\l}_{m,n} = - \left( {2\pi \over \tau_2}
\right)^2 ( w + m + n {\tau} ) ( - {\wb} + m + n {\bar\tau}),
\qquad w = i {\l} ( L_1 {\tau} - L_2 )} It is clear that the
modular group $SL_2 ({\IZ})$ permutes the eigenvalues ${\l}_{m,n}$
belonging to the different winding sectors. Now, we see that if
the modular parameter $\tau$ is an $SL_{2} ({\IZ})$ transform of
$i {\l} {p \over q}$, for $p, q \in {\IZ}_{+}$, $$ {\tau} = {a
{\tau}_0 + b \over c {\tau}_0 + d}$$then among the eigenvalues
${\l}_{m,n}$ there will be a zero, for the winding sector $(L_1,
L_2) = ( a p, c p)$. Conversely, for fixed $N = gcd (L_1, L_2)$
there will be a finite number of points ${\tau}_0$ in the
fundamental domain $\vert \tau \vert \geq 1, \ \vert \tau_1 \vert
\leq {1\over 2}$ of $SL_{2}({\IZ})$ such that the Laplacian will
have a zero mode obeying \bndryc\foot{Notice that for the
Euclidean orbifold, by the rotation $X^{+} = {\overline{X^{-}}}
\mapsto e^{2\pi i \l} X^{+}$, the analogous phenomenon occurs only
for rational ${\l}$, where for $N$ which is divisible by the
denominator of ${\l}$ the Laplacian will have zero modes. However
in that case $(L_1, L_2)$ should be defined modulo this
denominator as the twisting phases entering \bndryc\ are periodic
in $L_1, L_2$.}

If $\tau$ is generic then the spectrum of the Laplacian does not
contain zero. The regularized determinant is given by the standard
formula: \eqn\rdt{{\rm Det} ( -{\Delta} ) = {\exp} -
{\zeta}^{\prime} (0), \qquad {\zeta}(s) = \sum_{m,n}
{1\over{{\l}^{s}_{m,n}}}} It is easily computed to be \eqn\rdtt{
{\rm Det}( - {\Delta})^{-1} = \left\vert {{\eta}({\tau}) \over
{{\vt}_{1} ( w | {\tau} ) }} \right\vert^{2} \ {\exp} \left(
{{2\pi w_1^2 }\over{{\tau}_2}} \right)}

\ndt$\underline{\rm Full \ partition \ function}$ We can now
assemble all the pieces to write down the partition function for
fixed $N$: \eqn\fxdn{{\CZ}_{N} ({\t}, {\bar\t}) = \sum_{gcd(p,q) =
1} {d{\tau} d {\bar\tau} \over{{\t}_2^{13} {\vert \eta ({\tau})
\vert^{42}}}} e^{ G (N w_{p,q} \vert \t)}} where \eqn\grns{G (z,
{\t}) = - {\rm log} \left\vert {{\vt}_{1} (z \vert {\t} ) \over
{{\vt}_{1}^{\prime} (0 \vert {\t})}} \right\vert^2 + 2{\pi} {z_1^2
\over {\t}_2}} is the bosonic Green's function on the torus.

\ndt${\underline{\rm CFT \ interpretation}}$ Now let us go back to
the ${\IR}^{1,1}/{\IZ}$ contribution, of a fixed $(M,L )$ winding
sector: \eqn\small{Z_{M,L} ({\t}, {\bar\t}) = \left\vert {{{\eta}
({\t})}\over{{\vt}_{1} ( i {\l}(M {\t} - L ) \vert {\t} )}}
\right\vert^2 e^{- 2\pi {\l}^2 M^2 {\t}_2}} It is easy to check
the following modular properties of \small: \eqn\mdlr{\eqalign{&
Z_{M,L}\left( - {1\over {\t}}, -{1\over{\bar\t}} \right) = Z_{L,
-M} ({\t}, {\bar\t}) \cr & Z_{M,L} ({\t}+1, {\bar\t}+1) = Z_{M,
L-M} ({\t}, {\bar\t})\cr}} We would like now to perform a check of
our analysis of the old covariant quantization. Consider the
(pre)Hilbert space ${\CH}_{M}$ obtained by quantization in the
$M$-th winding sector. Let $g$ be the generator of the orbifold
group ${\IZ}$ acting in ${\CH}_{M}$.  We expect:
\eqn\trc{{\Tr}_{{\CH}_{M}} \ g^L \ q^{L_0 - {c\over 24}} {\bar
q}^{{\bar L}_{0} - {c\over 24}} =^{\kern -6pt ?}  Z_{M,L} ({\t},
{\bar\t})} Let us represent the Hilbert space as a tensor product
of the left and right-moving spaces. We see from \small\ that the
spectrum of $L_0$ and ${\bar L}_0$ is given by:
\eqn\spctrm{\eqalign{& L_0 = {({\l}M)^2 \over 2}  + ( N_{0} +
{\half} ) i {\l} M +  \sum_{n > 0} N^{+}_{n} (n + i{\l}M) +
N^{-}_{n} (n-i{\l}M) \cr  & {\bar L}_0 = {({\l} M)^2 \over 2} + (
{\bar N}_{0} + {\half} ) i {\l}M +  \sum_{n > 0} {\bar N}^{+}_{n}
( n - i {\l}M ) + {\bar N}^{-}_{n} ( n + i {\l}M ) \cr}} The range
of $N_0$ and ${\bar N}_0$ is slightly ambiguous: one can take $N_0
\geq 0$ or $N_0 < 0$ and similarly for ${\bar N}_0$. It actually
depends on $L$ what branch to choose (it has to do with the
spectral flow).

\subsec{Absence of the tachyon}

The major consequence of this analysis is that the twisted sector
contains no physical states. Indeed, the spectrum of $L_0$ and
${\bar L}_0$ is never real (the imaginary part is ${\l}$ times
half-integer), therefore the Virasoro constraints $L_0 = {\bar
L}_0 = 1$ will have no solutions. In the analogous case of
Euclidean rotational orbifold ${\IR}^2 / e^{2\pi i {\l}} $, the
Virasoro generators differ from \virs\ by ${\l} \to i {\l}$ and
one finds a physical spectrum, which turns out to contain a
tachyon (even in the case of superstring).

This conclusion about the absence of the physical states remains
correct in the case of superstring. Indeed, consider the NSR
string on this orbifold background and perform the old covariant
quantization \gsw. The tachyon can only come from the NS sector.
The twisted sector now in addition to the bosonic fields obeying
\tws\ contains the left and right-moving fermions ${\psi}_{L,R}$
obeying: \eqn\twsf{{\psi}_{L,R}^{\pm} ({\s} + {\pi} ) = - e^{\pm
2\pi M \l} {\psi}^{\pm}_{L,R} ({\s})} which amounts to the mode
expansion (we set $M=1$ for simplicity): \eqn\mdef{\eqalign{&
{\psi}_{L}^{\pm} = {1\over{{2}}} \sum_{r \in {\IZ} + {\half}}
{\tilde b}_{r}^{\pm} e^{ 2 ( \pm {\l} - i r ) ( {\t} + {\s})} \cr
& {\psi}_{R}^{\pm} = {1\over{2}} \sum_{r \in {\IZ} + {\half}}
b_{r}^{\pm} e^{ - 2 ( \pm {\l} + i r ) ( {\t} - {\s})} \cr }} The
components $b_{r}, {\tilde b}_{r}$ have the following reality
constraints: \eqn\relf{(b^{\pm}_{r})^{\dagger} = b_{-r}^{\pm},
\qquad ({\tilde b}^{\pm}_{r})^{\dagger} = {\tilde b}^{\pm}_{-r}}
and the canonical anti-commutation relations: \eqn\antic{ \{
b_{r}^{+}, b_{s}^{-} \} = - {\d}_{r+s} , \qquad \{ {\tilde
b}_{r}^{+} , {\tilde b}_{s}^{-} \} = - {\d}_{r+s}  } The ${\l}^2$
shift in the $L_0$ generator of Virasoro algebra now cancels
between bosons and fermions. However the imaginary part of $L_0$
is not cancelled: \eqn\nmbrs{ L_0 = L_0^{bos} - {{\l}^2 \over 2} +
\sum_{r \geq {\half}} - ( r + i \l ) b_{-r}^{-} b_{r}^{+} - ( - r
+ i \l ) b_{- r}^{+} b_{r}^{-} + r {\vec b}_{-r} \cdot {\vec
b}_{r}}

\ndt$\underline{\rm One-loop \ superstring \ measure}$ The type
IIA GSO projected one-loop measure is given by (after unfolding):
\eqn\sstr{\eqalign{ & A = \sum_{L=1}^{\infty} Z ( i{\l} L) \cr & Z
(u) =  \int {d^2 {\tau} \over { {\tau}_2^{5} }} {\vert {\vt}_{00}
(u) {\vt}^3_{00} (0) - {\vt}_{01} (u) {\vt}_{01}^3 (0) -
{\vt}_{10} (u) {\vt}_{10}^3 (0) \vert^2 \over \vert {\vt}_{11}(u)
{\eta}^9 ({\tau}) \vert^2 } = \cr & \qquad\qquad\qquad \int {d^2
\tau \over {\tau}_2^5 } \left\vert {{\vt}_{11}^4 ( u/2) \over
{{\vt}_{11} (u) {\vt}_{11}^{\prime 3} (0)}} \right\vert^2 \cr}} It
has poles at ${\tau}_{w,r, L} = {i {\l} L - r \over w}$ where
either $w$ or $r$ is not even (otherwise the pole from the
denominator is cancelled by the zero of the numerator). To derive
\sstr\ it is recommended to consult p. 33, vol. 2 of \joe, and to
use some Riemann identities.

\subsec{Stability}

It seems that the absence of the tachyon implies the stability of
the background. However one should keep in mind the example of the
open string in the electric field, where the absence of physical
states coexists with Schwinger instability having to do with the
creation of the charged open strings. The signal of this
instability -- the poles in the one-loop measure --- is currently
under investigation \costas.

\newsec{Orbifolds and quantum group}

In this section we discuss probing the orbifold geometry by
D-branes.  This topic is rather well-studied, however we believe
that our observation is new. We shall consider specifically
two-dimensional orbifolds.

First, let us look at the Euclidean orbifold: ${\IR}^2 / {\Gamma}$
where ${\Gamma}$ is a discrete subgroup of the rotation group
$SO(2)$. We shall assume that ${\Gamma}$ is generated by the
rotation by an angle $2{\pi} {\l}$. If ${\l}$ is rational then the
group ${\Gamma}$ is finite and the quotient is a cone. Let $q =
e^{2\pi i \l}$.

Now consider placing a D-particle on this quotient (times
otherwise flat space-time ${\IR}^{1,7}$). As usual, it lifts to a
${\Gamma}$-orbit on the covering space. The gauge theory supported
by the D-particle worldvolume is a truncation of the maximally
supersymmetric $U\left( \vert {\Gamma} \vert \right)$. We should
be thinking of ${\CH} = {\IC}^{\Gamma}$ -- the space of functions
on the orbit, as of the Chan-Paton space. The operation of
permuting the pre-images of the D-particle according to the
$\Gamma$-action leads to the unitary representation of ${\Gamma}$
in ${\CH}$ which is the regular representation. We shall consider
only regular branes. For these branes the truncation is simple.
Namely, the adjoint scalars $Z, Z^{*}$ corresponding to the motion
along the two-dimensional space we are taking quotient of, and the
adjoints $X^i$ corresponding to the rest of the space, must obey:
\eqn\orbc{\eqalign{& U^{-1} Z U = q Z \cr & U^{-1} Z^{*} U =
q^{-1} Z^{*} \cr & U^{-1} X^i U = X^i \cr}} where $U$ is some
unitary operator, representing the action of the rotation by
$2\pi\l$ in the Chan-Paton space. These conditions imply that the
spectrum of $Z, Z^{*}$ is ${\Gamma}$-invariant. The gauge group is
the subgroup of $U(\vert \Gamma \vert)$ which commutes with $U$.

The gauge theory action contains a potential $[Z , Z^{*}]^2$. In
the vacuum $[Z, Z^{*} ] =0 $ (technically speaking we should be
taking Dp-branes with $p > 1$ to speak about well-localized vacua
--- we shall keep this point in mind but will be speaking about
D-particles for simplicity). Thus, the vacua are the
representations of the following algebra: \eqn\algr{\eqalign{&
U^{-1} Z U = q Z \cr & U^{-1} Z^{*} U = q^{-1} Z^{*} \cr & [ Z,
Z^{*} ] = 0\cr}} It is obvious that the irreducible
representations are labelled by the points on the quotient space.
Indeed, as $Z$ and $Z^{*}$ commute, they can be diagonalized
simultaneously. Let $z$ be the eigenvalue of $Z$. Then $q z , q^2
z, \ldots$ are also eigen-values of $Z$.Hence in the irreducible
representation the spectrum of $Z$ coincides with an orbit of
${\Gamma}$. Moreover, replacing $z$ by $q^l z$ can be undone by a
gauge transformation with $U^l$. Hence the irreps up to an
isomorphism are labelled by the orbits.

We can even calculate the metric on the quotient space in terms of
the representation theory of the algebra \algr.  Let us write down
the kinetic term of the D-particle Lagrangian: $$ {\Tr} {\nabla} Z
{\nabla} Z^{*}, $$where ${\nabla} Z = {\dot Z} + [ A_0 , Z]$, and
integrate out $A_0$. We shall get the kinetic term $$ g_{z{\zb}}
{\dot z} {\dot {\zb}}$$ where $(z, {\zb})$ are some coordinates on
the moduli space of irreps of \algr, i.e. on the quotient space,
and $g_{z\zb}$ is the metric on the quotient.

Now let us give vev to some of the twisted sector fields. As we
discussed in the previous section there are physical states in the
twisted sector. As was shown in \douglasmoore\ the leading effect
(coming from the disk amplitude with one insertion of the twisted
sector field at the center and two open string vertex operators at
the boundary) of the NSNS sector twisted states is to replace the
potential $[Z, Z^{*}]^2$ by $\left( [Z, Z^{*}] - \sum_l {\m}_l U^l
\right)^2$. Here ${\m}_l$ is the vev of the tachyon in the $l$'th
twisted sector (corresponding to the rotation by $2{\pi}l {\l}$).
In this sum one can in principle add the $l=0$ term. This term
corresponds to turning on the constant $B$-field along the $Z$
plane, and does not come from the twisted sector. For simplicity
we shall assume that ${\m}_0 = 0$. The vacua of the D-particle
gauge theory now get deformed to: \eqn\dfalg{\eqalign{& U^{-1} Z U
= q Z \cr & U^{-1} Z^{*} U = q^{-1} Z^{*} \cr & [ Z, Z^{*} ] =
\sum_{l} {\m}_l U^l \cr}} Let us denote the algebra defined by
\dfalg\ by ${\CA}_{q, \m}$. The irreps of \dfalg\ are still
labelled by points of a two-dimensional manifold. And the same
kinetic term technique gives a metric on this manifold. This
metric now depends on $\m$ and is generically smooth. Here is a
quick way of calculating it.  Let us work in the basis in ${\CH}$
where $U$ is diagonal. Let $e^{i{\th}}$ be the eigenvalue of $U$.
Of course, it is easy to show that ${\th} = 2{\pi} {\l} l$, $l \in
{\IZ}$ but we shall not use this. It follows from \dfalg\ that
\eqn\nrmlfrm{ Z = e^{\pi {\l} {\p}_{\th}}  z ({\th}) e^{\pi \l
{\p}_{\th}}, \qquad Z^{*} = e^{-\pi \l {\p}_{\th}} {\zb} ({\th})
e^{-\pi \l {\p}_{\th}}} and that: \eqn\dff{z ({\th} + {\pi} {\l} )
{\zb} ({\th} + {\pi} {\l}) - z ({\th} - {\pi} {\l}) {\zb} ({\th} -
{\pi} {\l}) = \sum_l {\m}_l e^{i l {\th}} } The product $g ({\th})
= z ({\th}) {\zb} ({\th})$ is gauge invariant. From \dff\ we find:
$$ g ({\th}) = g + r ({\th}), \qquad r ({\th}) = \sum_{l} {{\m}_l
\over q^{l/2} - q^{-l/2} } e^{i l {\th}} $$ By the gauge
transformation commuting with $U$ we can bring $Z, Z^{*}$ to the
form \nrmlfrm\ with $z ({\th}) = \sqrt{g ({\th}} e^{i {\vf}},
{\zb} ({\th}) = \sqrt{g({\th})} e^{-i {\vf}}$. Then the metric on
the moduli space, parametrized by $g, {\vf}$ is calculated to be
\eqn\mtrc{ {dg^2 \over 4g_{eff}} + g_{eff} d{\vf}^2, \qquad
{1\over g_{eff}} = \sum_{\th} {1\over g({\th})}} If ${\l} =
{1\over N}$ then ${\th} = {k \over  N}, \ k = 0, \ldots , N-1$. At
the orbifold point, ${\m}_l = 0$, $g_{eff} = {g \over N}$, and the
metric \mtrc\ has ${\IZ}_{N}$ singularity at $g=0$. If ${\m}_l$
are generic then $g_{eff} \to 0$ at $g \neq 0$, namely at $g =
g_{0} = - {\rm min}_{\th} r ({\th})$, and locally the metric
\mtrc\ will look like $$ {dg^2 \over 4(g - g_0)} + ( g - g_0 )
d{\vf}^2 = dr^2 + r^2 d{\vf}^2$$for $r^2 = g - g_{0}$. If there
are several $\th$ for which $r({\th})$ reaches minimum then the
singularity will be of ${\IZ}_{N^{\prime}}$ type where
$N^{\prime}$ is the multiplicity of the minimum. The main
conclusion here is that turning on $\m$'s one smoothes out the
singularity. This conclusion is confirmed by other analysis
\aps\hmk.

Notice, that for a specific choice of ${\m}$ one can get a famous
algebra as the algebra of functions on the resolved quotient.
Namely, let us take $${\m}_l = {\m} { {{\d}_{l,1} - {\d}_{l,
-1}}\over{q - q^{-1}}}, \quad {\m} \in {\IR}_{+} $$ Introduce an
operator $H$, such that $U = q^{H}$, and the operators $E$ and
$F$, by the formulae: $E = Z / \sqrt{\m}$, $F = Z^{*} /
\sqrt{\m}$. Then the commutation relations \dfalg\ become that of
the quantum group $U_{q} sl_2$ ! Moreover the reality conditions
are those of the quantum $SU_q (2)$. For finite ${\Gamma}$ we need
$q$ to be the root of unity. This is the famous special case of
the quantum group representation theory.

One can ask at this point, what about $SL(2)$? In that case the
reality conditions would require $E$ and $F$ to be self-adjoint,
and to have unitary representations one would need $q$ to be real.
But this is exactly the setup for the Lorentzian orbifold we
started with!

Now, let us assume for a moment that the Lorentzian orbifold
twisted sector had physical states\foot{This would have been  the
case had the zero modes $\a_0^{\pm}$ been quantized in a unitary
manner, i.e. ${\a}_0^{-} \sim x, {\a}_0^{+} \sim i {\l} {\p}_x$.
In this case for the type IIA string  one finds a physical
spectrum in the sector with any winding number}. Let us also
assume for a moment that the slowly moving D-particle in the
time-dependent background such as \mln\ is described by the
Lorentzian version of the IKKT model. Then instead of static vacua
we can talk about the solutions to the matrix equations of motion
describing the (noncommutative) algebra of functions on the
D-brane worldvolume (as opposed to the usually considered algebra
of functions on its space-like section). We can now look at the
Lorentzian version of \dfalg\ and investigate the metric on the
moduli space.

One new feature compared to the ${\Gamma} = {\IZ}_N$ case is that
${\Gamma} = {\IZ}$ and the spectrum of ${\th}$ is continuous, in
fact ${\th} \in {\IR}/ 2\pi \IZ$. The formula \mtrc\ for the
metric is now changed to: \eqn\lmtrc{ {dg^2 \over 4 g_{eff} } -
g_{eff} d{\vf}^2 } where now: $$ {1 \over g_{eff}} = {1\over 2\pi}
\int {\rm d} {\th} {1\over g({\th})}, \qquad g ({\th}) = g +
\sum_{l} {{m}_l \over {\rm sinh} ({\pi} l {\l})} e^{i l {\th}} $$
Now for $g$ near the minimum of $r({\th})$ the behaviour of
$g_{eff}$ is different: $$ g_{eff} \sim \sqrt{ g - g_{0}}$$ and
the metric \lmtrc\ looks more singular then the original orbifold
metric!

It would have been a very bizarre situation if by turning on the
twisted sector fields one made the singularity worse then it was
before! Luckily there are no twisted sector fields to turn on, and
no contradiction.

\centerline{\bf Acknowledgements}

I would like to thank G.~Moore, N.~Seiberg for their interest and
discussions, C.~Bachas for collaboration in \costas, and
A.~Gorsky, I.~Kogan, A.~Losev,  A.~Morozov,  and B.~Pioline for
useful discussions.

I would like to thank the organizers of the December 2001 workshop
at Annecy and the organizers of ITEP winter school for warm
hospitality and for invitation to present some of the results
above. I am also grateful to the New High Energy Theory Center at
Rutgers University for its kind invitation in October 2001 which
helped me to perform this study. I am especially grateful to my
family for its support and understanding.

Research was partially supported by RFFI under grant 00-02-16530
and under the grant 00-15-96557 for scientific schools, and by
Clay Mathematical Institute.

\ndt{\bf Note added.} After this paper has been completed we have
received the manuscripts \cornalba\vijay\sb\ which address
somewhat related issues. In particular, the paper \cornalba\ also
analyzes the bosonic string one-loop partition function in the
Lorentzian orbifold background.

\footatend\vfill\supereject\immediate\closeout\rfile\writestoppt
\baselineskip=14pt\centerline{{\bf References}}\bigskip{\frenchspacing%
\parindent=20pt\escapechar=` \input refs.tmp\vfill\eject}\nonfrenchspacing \bye